\documentclass[%
twocolumn,amssymb, amsmath,nofootinbib,showpacs, aps, prl]{revtex4}
\usepackage{docs}%
\usepackage{bm}%

\expandafter\ifx\csname package@font\endcsname\relax\else
 \expandafter\expandafter
 \expandafter\usepackage
 \expandafter\expandafter
 \expandafter{\csname package@font\endcsname}%
\fi

\begin{document}

\title{Constraints on the Quark Correlation Matrix \\
from Equations of Motion}
\author{Elvio Di Salvo}%
\email{disalvo@ge.infn.it}
\affiliation{{Dipartimento di Fisica and I.N.F.N. - Sez. Genova, Via Dodecaneso, 33 \\-
 16146 Genova, Italy\\}}
\date{July 2004}%

\begin{abstract}
We show that the Politzer theorem on the equations of motion implies strong
constraints on the quark correlation matrix, restricting the number of 
independent distribution functions that characterize the internal structure of 
the nucleon. Then we draw some important consequences from this result. First of 
all, we suggest an alternative method for determining transversity. Secondly, we 
predict the $Q^2$-dependence of some azimuthal asymmetries. Lastly, we illustrate
the origin of the main contributions to some twist-3 distribution functions.
\end{abstract}


\maketitle

$~~~~$The  theoretical expressions for inclusive reaction cross sections at high 
momentum transfers are typically of the form\cite{co1} 
\begin{equation}
d\sigma \propto {\cal S}_1 \otimes {\cal S}_2 \otimes .... \otimes {\cal S}_m 
\otimes {\cal H}_1 \otimes {\cal H}_2 \otimes .... \otimes {\cal 
H}_n, \label{cs}   
\end{equation}
where the symbol "$\otimes$" means convolution and/or matrix product. The ${\cal 
S}_k$ ($k$ = 1, ... m) denote "soft" functions, describing either the internal structure of 
the initial hadrons participating in the reaction, or the parton fragmentation into 
the final detected hadrons. On the other hand, the ${\cal H}_l$ ($l$ = 1, ... n) describe 
the "hard" scattering, where the active partons are involved. Eq. (\ref{cs}) is known as the 
factorization property. As regards semi-inclusive reactions - which are receiving an 
increasing attention from spin physicists - this property was proven many years ago for 
back-to-back hadron pair production in $e^+-e^-$ collisions\cite{cs1} and recently also 
for semi-inclusive deep inelastic scattering (SIDIS) and Drell-Yan\cite{jmy}, which involve 
tranverse momentum dependent (tmd) "soft" functions. Moreover one sometimes adopts the 
handbag\cite{mt,co2} approximation, neglecting\cite{jj}, either the interactions between 
such partons and the spectator ones after the hard scattering process\cite{bhs} (in 
reactions 
like deep inelastic), or between the active quark and the spectator partons of other hadrons 
involved in the reactions ({\it e. g.}, in Drell-Yan). In this approximation the "soft" 
factors may be encoded in the so-called correlation and/or fragmentation matrix\cite{mt}.      

Mulders and Tangerman\cite{mt} (MT) used the Politzer theorem\cite{po}, concerning the 
equations of motion, to establish constraints between quark-quark and 
quark-quark-gluon correlation matrices.  Here we draw from that theorem more 
dramatic consequences on the tmd quark-quark correlation 
matrix, showing that, in the framework of QCD, it fulfils an inhomogeneous Dirac 
equation. This strongly restricts the number of independent elements
in that matrix, while uniquely fixing a given energy scale that appears in the 
parametrization of some of the "soft" functions. In turn this suggests, among other things, 
an alternative, promising method for extracting the transversity function from data. 
Moreover, the result allows to predict the $Q^2$-dependence for azimuthal asymmetries[7-11].
Lastly, a new light is cast on some twist-3 functions, generalizing a result about 
$g_2(x)$\cite{et}.  

We denote the correlation matrix by $\Phi$, which we define as
\begin{eqnarray}
&~&\Phi_{ij}(p; P,S) = \nonumber 
\\
&~&\int\frac{d^4x}{(2\pi)^4} e^{ipx} 
\langle P,S|\bar{\psi}_j(0) {\cal L}(x)  \psi_i(x)|P,S\rangle. \label{corr}
\end{eqnarray}
Here $|P,S\rangle$ is a nucleon state with a given four-momentum $P$ and Pauli-Lubanski (PL) 
four-vector $S$. Moreover
\begin{equation}
{\cal L}(x) = {\mathrm P} exp\left(-ig\int_0^x 
\lambda_a A^a_{\mu}(z)dz^{\mu}\right) \label{link}
\end{equation}
is the gauge link operator\cite{mt}, $g$, $\lambda_a$ and $A^a_{\mu}$ being respectively the 
strong coupling constant, the Gell-Mann matrices and the gluon fields. The link operator 
depends on the choice of the integration path, say ${\cal P}$, which has to be fixed so as 
to make a physical sense\cite{col,cs,co2}. According to previous treatments\cite{mt,cs}, we 
define the path - to be denoted by ${\cal P}_+$ - as a set of three pieces of straight line, 
from the origin to $x_{1\infty}\equiv (+\infty, 0, {\bf 0}_{\perp})$, from $x_{1\infty}$ to 
$x_{2\infty}\equiv (+\infty, x^+, {\bf x}_{\perp})$ and from $x_{2\infty}$ to  $x\equiv 
(x^-, x^+,{\bf x}_{\perp})$. Here we have adopted a frame - to be used throughout this 
letter - whose $z$-axis is taken along the nucleon momentum; moreover we have introduced 
light cone coordinates. 

Two remarks are in order. Firstly, we have omitted the flavor and color indices of the 
quark, according to the MT notation; similarly, we shall drop the flavor index from the 
"soft" functions. Secondly, for T-odd functions\cite{bm}, the path ${\cal P}_+$ is suitable, 
for example, for SIDIS, while for Drell-Yan one has to consider the one through $x^- = 
-\infty$\cite{col,jy,bjy,bmp} to be denoted by ${\cal P}_-$.

Now we deduce an equation for  the correlation matrix in the framework of QCD. 
To this end, first of all, we recall the Politzer theorem\cite{po}, {\it i. e.},  
\begin{equation}
\langle h|{\cal F}(\psi) (i\rlap/D-m_q)\psi(x)|h\rangle = 0. \label{pol}
\end{equation}
Here $m_q$ is the quark rest mass, $|h\rangle$ a hadronic state, ${\cal F}(\psi)$
a functional of the quark field and $D_{\mu}$ the covariant derivative,
$D_{\mu} = \partial_{\mu} - ig \lambda_a A^a_{\mu}$.
The result (\ref{pol}) survives renormalization. Furthermore, taking into account
the path ${\cal P}_+$ in ${\cal L}(x)$, the covariant derivative implies
\begin{eqnarray}
&~&{\cal L}(x)(i\rlap/D-m_q)\psi(x) = (i\rlap/\partial -m_q) 
\left[{\cal L}(x)\psi(x)\right] \nonumber 
\\
&~&+  g\lambda_{a} \rlap/B_{\perp}^a(x) {\cal L}(x) \psi(x). \label{r1}
\end{eqnarray}
Here
\begin{equation} 
B_{\perp}^{a\mu}(x) = \int_{x^-}^{\infty} dy^- H_-^{a\mu}(y^-, x^+,  
{\bf x}_{\perp}) \label{rr}
\end{equation}
and $H_-^{a\mu} =\partial_- A^{a\mu} - \partial^{\mu} A^a_-.$
From eqs. (\ref{pol}) and (\ref{r1}) - setting ${\cal F}(\psi)$ = ${\bar \psi}(0) {\cal 
L}(x)$ and $|h\rangle$ = $|P,S\rangle$ - we get 
\begin{equation}
\langle P,S|{\bar \psi}_j(0)(i\rlap/\partial-m_q)\left[{\cal L}(x)\psi_i(x)
\right]|P,S\rangle = g\beta_{ij}, \label{r2}
\end{equation}
where
\begin{eqnarray}
\beta_{ij} &=& \beta_{ij}(x; P,S) \nonumber \\
 &=& -\langle P,S|{\bar \psi}_j(0) \lambda_a \rlap/B_{\perp}^a(x) 
{\cal L}(x) \psi_i(x)|P,S \rangle. \label{r12}
\end{eqnarray}
Moreover
\begin{equation}
\int d^4x e^{ipx}\rlap/\partial\left[{\cal L}(x)\psi(x)\right]
= -i \rlap/p\int d^4x e^{ipx}{\cal L}(x)\psi(x), \label{r3}
\end{equation}
since $\psi(x)$ must vanish at infinite distances, in order to guarantee 
four-momentum conservation. Fourier transforming both sides of eq. 
(\ref{r2}), and applying the result (\ref{r3}), we get
\begin{equation}
(\rlap/p - m_q)\Phi (p; P,S) = g {\tilde \beta} (p; P,S),\label{dirac11}
\end{equation}
where 
\begin{equation}
{\tilde \beta}(p; P,S) = \int\frac{d^4x}{(2\pi)^4} e^{ipx} \beta (x; P,S). 
\label{inho} 
\end{equation}
and the $\beta$ matrix is defined through eq. (\ref{r12}).
The solution to eq. (\ref{dirac11}) is of the type
\begin{equation}
\Phi (p; P,S) = (\rlap/p + m_q)\Psi,\label{matc}
\end{equation}
where
\begin{eqnarray}
&~&\Psi = \delta(p^2 - m_q^2) \Psi_f (p; P,S)+\nonumber
\\
&~&g{\tilde \beta}(p; P,S)(p^2-m_q^2-i\epsilon)^{-1}.\label{matr}
\end{eqnarray}
Here the first term of eq. (\ref{matr}) corresponds to the correlation matrix of an on-shell 
quark in the QCD improved parton model. On the contrary, the second term takes into account 
quark offshellness and interaction dependence. Hermiticity, Lorentz invariance and  parity 
invariance for $\Phi$ bind $\Psi$ to be a linear combination of 5 Dirac operators, {\it i. 
e.},  
\begin{equation}
\Psi = \sum_{i=1}^5 \Gamma_i B_i(p^2, P\cdot p), \label{param}
\end{equation}
where $B_i$ are Lorentz invariant real functions, while 
\begin{eqnarray}
\Gamma_1 &=& 1, ~~~~ \Gamma_2 = \gamma_5 \rlap/S_{q\parallel}, ~~~~ \Gamma_3 = \gamma_5 
\rlap/S_{q\perp}, \nonumber
\\ 
\Gamma_4 &=& -i \rlap/P_{\perp}/\mu, ~~~~~~~ \Gamma_5 = -i \rlap/P_{\perp}\gamma_5 
\rlap/S_q/\mu. \nonumber
\end{eqnarray} 
Here $S_q$ is the PL vector of the active quark,
\begin{eqnarray} 
S_{q\parallel} &=& (S_q\cdot n_+) n_-+(S_q\cdot n_-) n_+, \nonumber \\
S_{q\perp} &=& S_q-S_{q\parallel} \nonumber
\end{eqnarray}
and $n_{\pm}$ are lightlike vectors, such that $n_+\cdot n_- = 1$ and whose space components 
are directed along the momentum of the quark. Moreover
\begin{equation}
P_{\perp} = P - (P\cdot n_+) n_- - (P\cdot n_-) n_+. \nonumber
\end{equation}
Lastly $\mu$ is a parameter with the dimensions of a momentum, introduced for dimensional 
reasons and to be determined below. 

Now we insert eq. (\ref{param}) into (\ref{matc}), observing that for $i$ = 1 to 3 the 
operators $\rlap/p\Gamma_i$ are twist 2 and therefore interaction independent, while for $i$ 
= 4 and 5 such operators are T-odd and therefore depend on interactions, since they describe 
interference effects\cite{le}. Therefore, comparing eq. (\ref{matr}) with eq. (\ref{param}) 
yields 
\begin{equation}
\Psi_f = \sum_{i=1}^3 \Gamma_i B'_i,  \ ~~~~ \ ~~~~~ \
g{\tilde \beta} = \sum_{i=4}^5 \Gamma_i B'_i,\label{int}
\end{equation}
where 
\begin{eqnarray}
B_i &=& B'_i\delta(p^2-m_q^2), ~~~~~~~~ i = 1, 2, 3, \nonumber \\
    &=& B'_i(p^2-m_q^2-i\epsilon)^{-1}, ~~~~ i = 4, 5. \nonumber
\end{eqnarray}
Notice that ${\tilde \beta}$ depends crucially on the path ${\cal P}$, see eqs. (\ref{r1}) 
and (\ref{rr}); in particular it changes sign according as to whether ${\cal P}$ = ${\cal 
P}_+$ or ${\cal P}_-$, as expected for T-odd functions\cite{col}. 

Now we introduce the projections, integrated over $p^-$, of the correlation matrix over the 
Dirac components\cite{mt}, {\it i. e.}, 
\begin{equation}
\Phi^{\Gamma}(p^+, {\bf p}_{\perp}; P,S) = 
\frac{1}{2}\int dp^- tr(\Gamma\Phi),\label{proj}
\end{equation}
where $\Gamma$ is a Dirac operator and $p \equiv (p^-, p^+, {\bf p}_{\perp})$. Inserting 
eqs. (\ref{matc}) and (\ref{matr}) into (\ref{proj}), and closing, according to eq. 
(\ref{inho}), the integration path in the upper complex half-plane, we get
\begin{eqnarray}
&~&\Phi^{\Gamma}(p^+, {\bf p}_{\perp}; P,S) = \nonumber
\\
&~&\frac{1}{4p^+} tr\left[\Gamma(\rlap/p_0+m_q)(\Psi_{f0}+2\pi
ig{\tilde\beta}_0)\right].\label{proj2}
\end{eqnarray}
Here $\Psi_{f0}$ and ${\tilde\beta}_0$ are, respectively, the functions $\Psi_f$ and 
${\tilde\beta}$ calculated at $p = p_0$, where
$p_0 \equiv [(m_q^2+{\bf p}^2_{\perp})/2p^+, p^+, {\bf p}_{\perp}]$. 
On the other hand, the projections over the Dirac components may be expressed
according to the usual notations of the tmd functions\cite{mt}. For example, we have
\begin{eqnarray}
\Phi^{\gamma^+} &=& f_1,\\
\Phi^{\gamma^5\gamma^+} &=&  \Lambda g_{1L} + \lambda_{\perp} g_{1T},\label{lg}\\ 
\Phi^{\gamma^5\gamma^l\gamma^+} &=& S^l h_{1T} + \pi_{\perp}^l \left(\Lambda h_{1L}^{\perp}+
\lambda_{\perp} h_{1T}^{\perp}\right), \label{trsv} \\
\Phi^{\gamma^l\gamma^+} &=& \pi_{\perp}^l h_{1}^{\perp}, \\
\Phi^{\gamma^in_{\perp i}} &=& \pi_{\perp}^in_i f_{1T}^{\perp}. 
\end{eqnarray}
Here $\Lambda = MS^+/P^+$ is the nucleon light cone helicity, $M$ the nucleon mass, 
$\lambda_{\perp} = -p_{\perp}\cdot S/\mu$ and $p_{\perp} = p-(P\cdot p/M^2)P$, in our frame 
$p_{\perp}\equiv (0, 0,{\bf p}_{\perp})$. Moreover $\pi_{\perp} = p_{\perp}/\mu$ and 
$n_{\perp}$ is such that $n_{\perp}^2$ = -1, $n_{\perp}\cdot P$ = $n_{\perp}\cdot p_{\perp}$ 
= 0. Comparing such relations with the projections (\ref{proj2}) along the same Dirac 
components, we obtain some important results. First of  all, we identify 
\begin{eqnarray}
f_1 &=& B'_{1,0}, ~~~~ g_{1L} = B'_{2,0},\nonumber\\  
h_{1T} &=& B'_{3,0}, ~~~~ h_{1}^{\perp} = 2\pi B'_{4,0},\nonumber \\
f_{1T}^{\perp} &=& 2\pi B'_{5,0},\nonumber
\end{eqnarray}
 where the $B'_{i,0}$ are the functions $B'_i$ calculated at $p = p_0$. It follows that the 
5 functions listed above - $f_1$, $g_{1L}$, $h_{1T}$, $h_{1}^{\perp}$ and $f_{1T}^{\perp}$ - 
are Lorentz invariant, and therefore they depend on the Bjorken variable $x$, on $Q^2$ and 
on $p_{\perp}^2$.  Incidentally, if we consider the correlation matrix (\ref{matc}) in the 
limit of $g \to 0$ and integrate both sides of that equation over $p^-$, we get (up to a 
normalization factor) the spin density matrix of a free, on-shell quark, {i. e.}, 
\begin{eqnarray}
\rho &=& 1/2(\rlap/p_0+m_q)\Psi_{f0}, \label{dm}
\\
\Psi_{f0} &=& f_1 + g_{1L}\gamma_5 \rlap/S_{q\parallel} + h_{1T}\gamma_5 
\rlap/S_{q\perp}\label{dm1}.
\end{eqnarray}
 Secondly we establish some relations among the traditional\cite{mt} "soft" functions, as 
expected, since our parametrization consists of 5 independent functions, instead of the 12 
found by MT. For example, from eqs. (\ref{lg}) and (\ref{trsv}), it follows 
\begin{equation}
g_{1T} = h^{\perp}_{1T}\frac{1-\epsilon_1}{1-\epsilon_2} = 
\frac{|{\bf p}|}{\mu}h_{1T}(1-\epsilon_1). \label{meq}
\end{equation}
Here ${\bf p}$ is the momentum of the quark. Moreover 
\begin{eqnarray}
1-\epsilon_1 &=& cos \theta - \alpha (1-cos\theta),
\\
1-\epsilon_2 &=& A - \alpha (A+1),
\end{eqnarray}
$A = \sqrt{2}xP^+/|{\bf p}| - cos \theta$, $sin\theta =  |{\bf p}_{\perp}|/|{\bf p}|$ and 
$\alpha = [(m_q^2+{\bf p}^2)^{1/2}-|{\bf p}|]/m_q$. 
In deriving result (\ref{meq}), we have taken into account that $S_q$ does not coincide with 
$S$: indeed, one has $S \equiv (0,{\bf S})$ in the {\it nucleon} rest frame and $S_q \equiv 
(0,\pm{\bf S})$ in the {\it quark} rest frame\cite{ael,dis1}, with ${\bf S}^2 = 1$.

Furthermore, $g_{1T}$ and $h^{\perp}_{1T}$ can be interpreted as probability densities, as 
well as $h_{1T}$, provided they are appropriately normalized. Therefore we have to fix
\begin{equation}
\mu = |{\bf p}|. \label{mu}
\end{equation}
This result was also found starting from a simple model\cite{dis2} for T-odd functions, 
based on the interference between two different partial waves\cite{bhs,bhs2,jm}. 

It is to be noticed that, by considering the projections (\ref{proj2}) over all the Dirac 
components, one obtains several relations of the type (\ref{meq}) among the usual T-even tmd 
functions\cite{mt}. Several such relations can be established by considering the  
projections over the Dirac components of eq. (\ref{dm}), which immediately shows that the 
independent T-even tmd functions amount to $f_1$, $g_{1L}$ and $h_{1T}$. All other 
T-even functions are related to these through frame dependent multiplicative factors, as can 
be seen, for example, from eqs. (\ref{meq}).

Each of the 5 independent functions characterizing the correlation matrix is related to an 
{\it u}-channel helicity amplitude for quark-nucleon elastic scattering\cite{bls,so}:
\begin{eqnarray}
&~&f_1 = \phi^u_1 +\phi^u_2, ~~~~~~~ g_{1L} = \phi^u_1 -\phi^u_2,\nonumber \\
&~&h_{1T} = \phi^u_3, ~~~~~~~~~~~~ sin\theta h_1^{\perp} = \phi^u_4,\nonumber \\
&~&sin\theta cos\varphi  f_1^{\perp} = \phi^u_5.\nonumber  
\end{eqnarray}
Here $cos\varphi = -p_{\perp}\cdot S/|{\bf p}_{\perp}|$ and the amplitudes $\phi^u_l$ ($l$ = 
1 to 5) are obtained by analytical continuation of the {\it s}-channel amplitudes
\begin{eqnarray}
\phi^s_1 &=& -i\langle ++|T-T^{\dagger}|++ \rangle, \nonumber \\ 
\phi^s_2 &=& -i\langle +-|T-T^{\dagger}|+- \rangle, \nonumber \\
\phi^s_3 &=& -i\langle +-|T-T^{\dagger}|-+ \rangle, \nonumber \\
\phi^s_4 &=& -i\langle ++|T-T^{\dagger}|+- \rangle, \nonumber \\
\phi^s_5 &=& -i\langle ++|T-T^{\dagger}|-+ \rangle. \nonumber
\end{eqnarray}
We have denoted by $T$ the T-matrix and by $|\Lambda, \lambda \rangle$ the helicity states,  
where $\Lambda$ and $\lambda$ are respectively the nucleon and quark helicity. Notice that 
pure symmetry considerations, independent of the details of the interaction,  would yield 6 
independent helicity amplitudes, since in this case only parity and time reversal invariance 
can be applied\cite{bls}; therefore we conclude that a further constraint is implied by 
gauge invariance, as we have assumed.

To conclude, we sketch some consequences of our theoretical results.

A) Eq. (\ref{trsv}) implies the following expression of the transversity:
\begin{equation} 
h_1(x) = \int d^2 p_{\perp} \left[h_{1T}(x,{\bf p}_{\perp}^2)+
 \lambda_{\perp}^2 h_{1T}^{\perp}(x,{\bf p}_{\perp}^2)\right]. \label{tr1} 
\end{equation} 
We point out that this expression contains a term - the second one of eq. (\ref{tr1}) - 
which is frame dependent, since $\lambda_{\perp} = sin\theta cos \varphi$ and 
$h_{1T}^{\perp}$ contains the factor $(1-\epsilon_2)$, see eq. (\ref{meq}). To illustrate 
this term, consider a transversely polarized nucleon. The quark longitudinal polarization, 
due in this case to the transverse momentum, is magnified by the boost from the quark rest 
frame. This additional, frame dependent, polarization has a component along the transverse 
momentum, see the second term  of eq. (\ref{trsv}). 
Furthermore eq. (\ref{tr1}), together with eqs. (\ref{meq}) to (\ref{mu}), suggests an 
alternative method for determining the transversity of a nucleon. Indeed, $g_{1T}$ can be 
conveniently extracted from double spin asymmetry\cite{km,dis1,dis3} in SIDIS with a 
transversely polarized target. This asymmetry is expressed as a convolution of the unknown 
function with the usual, well-known fragmentation function of the pion. Therefore the method 
appears more convenient than the  usually proposed one\cite{her}, based on the Collins 
effect\cite{co4} in single spin SIDIS asymmetry, since in this case one is faced with a  
convolutive product of two unknown functions, $h_{1T}$ and the Collins function.   

B) Eq. (\ref{mu}) determines the $Q^2$-dependence of the azimuthal asymmetries whose 
theoretical expressions in the formalism of the correlation matrix contain the factor 
$\mu^{-1}$. Typical cases are the two above mentioned SIDIS asymmetries, for which we 
predict a decrease like $Q^{-1}$, contrary to the current literature\cite{mt}. Our result 
follows
immediately from eq. (\ref{mu}), by calculating the quark momentum in the Breit frame where 
the the time component of the virtual photon momentum vanishes. Analogously, we predict a 
decrease like $Q^{-2}$ for the polarized SIDIS and Drell-Yan azimuthal asymmetries, which, 
in the formalism of the correlation matrix, result in convolutive  products of two T-odd 
functions\cite{bjm}. In particular, in the case of Drell-Yan, data\cite{fal} exhibit such an 
azimuthal asymmetry, which, at least for ${\bf p}_{\perp}^2 << Q^2$, can be conveniently 
interpreted in the framework of that formalism\cite{bbh}; but the $Q^2$ dependence  is 
consistent with our prediction\cite{dis4} and not with the MT statement, $\mu = M$.  

C) Eqs. (\ref{matc}) and (\ref{matr}) imply that, in the handbag approximation, the T-even 
twist 3 "soft" functions\cite{mt} are proportional either to $|{\bf p}_{\perp}|$ or to 
$m_q$. As a consequence of this fact, the most significant contributions to the ${\bf 
p}_{\perp}$-integrated  functions $g_T = g_1+g_2$, $h_L$ and $e_1$ come from non-handbag 
diagrams,  generalizing the result of Efremov and Teryaev for $g_2$\cite{et,bdr}. 
\vskip 0.30in

\centerline{\bf Acknowledgements}
The author is grateful to his friends A. Blasi, A. Di Giacomo and M. Pusterla for fruitful 
discussions.

\end{document}